\documentclass[aps,groupedaddress, nofootinbib, twocolumn, showkeys]{revtex4-2}
\usepackage{graphicx}
\usepackage{amsmath}
\usepackage{graphicx}

\usepackage[caption=false]{subfig}
\usepackage{dcolumn}
\usepackage{bm}
\usepackage{siunitx}
\usepackage{mathtools}
\usepackage[table]{xcolor}
\usepackage{float}
\usepackage{booktabs}
\usepackage{multirow}

\usepackage[colorlinks=true, linkcolor=blue, citecolor=blue, urlcolor=blue]{hyperref}

\def \spintec {Univ. Grenoble Alpes, CEA, CNRS, Grenoble INP, SPINTEC, Grenoble, France}
\def \cnrs {Centre de Nanosciences et de Nanotechnologies, CNRS, Université Paris-Saclay, Palaiseau, France}

\begin{document}
\raggedbottom 

\title{State Diagram of Dual-Free-Layer Perpendicular Magnetic Tunnel Junctions for Artificial Spiking Neurons}%

\author{Arthur Courberand\textsuperscript{1, $\ddagger$}}
\email{arthur.courberand@cea.fr}
\author{Ajay Jha\textsuperscript{1, $\ddagger$}}
\email{ajay-kumar.jha@cea.fr}
\author{Ricardo C. Sousa$^{1}$}
\author{Stephane Auffret$^{1}$}
\author{Joo-Von Kim$^{2}$}
\author{Ursula Ebels$^{1}$}
\author{Liliana D. Buda-Prejbeanu$^{1}$}
\affiliation{$^1$\spintec, $^2$\cnrs }

\begin{abstract}
The performance of perpendicular magnetic tunnel junctions (MTJs) depends critically on interfacial properties and layer thicknesses, small variations of which can substantially modify their magnetic and transport characteristics. This work investigates, experimentally and numerically, a perpendicular MTJ with two free layers, a configuration that preserves the precessional dynamics of the device while simplifying its fabrication.
Quasi-static state diagrams measured under magnetic fields and DC or pulsed bias reveal static parallel, antiparallel, and bistable regimes, as well as a dynamical \emph{windmill} regime in which the reversal of one layer triggers that of the other, giving rise to spiking neuron-like signals.
Macrospin simulations reproduce these diagrams, and a stability analysis of the Landau-Lifchitz-Gilbert (LLG) equation yields an analytical description of the critical lines, including the \emph{windmill} threshold voltage. The comparison also reveals a quadratic term in the voltage dependence of the spin-transfer torque that Joule heating cannot explain. The analytical model shows that the threshold voltage can be lowered either by approaching the spin-reorientation transition, at the cost of thermal stability, or by enhancing the quadratic contribution to the spin-transfer torque. These results establish guidelines for optimizing multilayer MTJ stacks for artificial spiking neurons and a broader range of neuromorphic and spintronic applications.
\end{abstract}

\maketitle
\def\thefootnote{$\ddagger$}\footnotetext{These authors contributed equally to this work}

\section{Introduction}

Magnetic tunnel junctions (MTJs) are spintronic devices consisting of two ferromagnetic layers separated by a thin insulating tunnel barrier~\cite{julliere1975tunneling, moodera1995large, butler2001spin}. They host a remarkably rich variety of magnetization dynamics, stochastic or deterministic switching, steady-state precession, domain wall nucleation, all of which can be read out directly through the device resistance. This electrical access is possible due to the tunnel magnetoresistance (TMR) effect, which ties the resistance to the relative orientation of the two magnetizations~\cite{julliere1975tunneling, moodera1995large}. Conversely, a current flowing through the junction becomes spin-polarized and exerts a spin-transfer torque (STT) on the magnetization, capable of driving it out of equilibrium or reversing it altogether~\cite{slonczewski1996current, berger1996emission}. MTJs are already widely used in practical applications: they are the core of MRAM memories~\cite{kent2015new, worledge2024spin}, form the read heads of hard disk drives~\cite{shiroishi2009future, zhu2006magnetic}, and underpin modern magnetic sensors~\cite{khan2021magnetic, zhu2006magnetic, jha2026low}.

Their spin-dependent transport and magnetization dynamics also inherently provide nonlinearity, hysteresis, and non-volatility at the nanoscale, which are key attributes for compact and energy-efficient neural hardware~\cite{cai2019voltage, locatelli2014spin}. In particular, an MTJ can act as a compact memristor~\cite{mansueto2021spintronic, locatelli2014spin}. Its non-volatile states and sharp switching threshold make the MTJ useful not only for storing synaptic weights, but also for mimicking neuronal integration and spiking~\cite{vincent2015spin, lequeux2016magnetic}. For instance, the magnetization dynamics of the free layer can implement a leaky integrate-and-fire (LIF) behavior, spin-current pulses are summed into the magnetic state, and once the total input surpasses the energy barrier, the MTJ switches and emits a spike~\cite{mansueto2021spintronic, locatelli2014spin, grollier2016spintronic}.

In conventional MTJs, one layer is pinned, usually by a synthetic antiferromagnet (SAF), and acts as a fixed spin polarizer, so that only the free layer responds to STT. In this configuration, the macrospin Landau-Lifshitz-Gilbert-Slonczewski (LLGS) equation can be treated analytically: a linear stability analysis yields the critical current for switching~\cite{slonczewski1996current, sun2000spin, katine2000current}, and the stability diagram in the current--field plane distinguishes static states, switching, and steady-state precession~\cite{bernert2014phase, timopheev2015respective}. At finite temperature, the same framework describes thermally activated switching over an effective energy barrier reduced by the current~\cite{li2004thermally, koch2004time}. These analytical descriptions have provided the design rules of STT devices, linking thermal stability, critical current, and switching time.

A particularly promising device for artificial neurons is the dual-free-layer MTJ (DFL-MTJ), in which both magnetic layers are free to switch. The dynamics of this system was first described by Matsumoto \textit{et al.}~\cite{matsumoto2019} which showed that under an applied DC bias, mutual spin-transfer torques induce a sustained \emph{windmill} dynamics, whereby the two layers reverse sequentially in an alternating manner. Each reversal toggles the junction between the parallel and antiparallel configurations, producing a sharp resistance transition that serves as an electrical spike, which yields continuous, repetitive firing. DFL-MTJs were then proposed as probabilistic bits~\cite{camsari2021double}. In our previous work DFL-MTJs were demonstrated experimentally as spiking neurons~\cite{farcis2023spiking}. Crucially, the firing rate is voltage-tunable, adjusting the bias smoothly changes the oscillation frequency from the kHz to the tens of MHz range. Moreover, the device operates at room temperature with no external field, and the energy per spike is only a few picojoules~\cite{farcis2023spiking}. In practice, the DFL-MTJ behaves as a very fast LIF-like neuron.

DFL-MTJs exhibit intrinsic stochastic behavior arising from thermally assisted spin-transfer torque, resulting in spontaneous telegraph-like resistance fluctuations across a broad bias range. This property enables their operation as probabilistic neurons (\emph{p-bits}), where the output spike probability can be continuously modulated by the applied voltage~\cite{camsari2021double}. The symmetric stack configuration of the DFL-MTJ eliminates the need for pinned/reference layers, thereby simplifying the device architecture while retaining compatibility with established MRAM fabrication processes. From a technological perspective, the industrial maturity of STT-MRAM provides a scalable, CMOS-compatible foundation for the development and large-scale integration of spintronic neuromorphic systems. Extending this framework toward analog and spiking MTJ-based crossbar arrays offers a viable pathway to hardware accelerators for spiking neural networks. DFL-MTJs are particularly compelling, as they integrate high-speed dynamics, voltage tunability, and controllable stochasticity within a compact nanoscale device suitable for energy-efficient brain-inspired computing architectures~\cite{mehonic2024roadmap}.

These properties make DFL-MTJs natural candidates for the hardware implementation of spiking neural networks (SNNs), which exploit sparse, temporal spike coding to reduce energy consumption compared with conventional artificial neural networks~\cite{yamazaki2022spiking, marrows2024neuromorphic, merolla2014million, davies2018loihi}. Their elementary unit is usually the LIF neuron, which integrates input currents into a leaky membrane potential and emits a spike when a threshold is reached~\cite{burkitt2006review, indiveri2011neuromorphic}. Several MTJ-based neurons have already been demonstrated, using multidomain free layers~\cite{wang2022noise} or current-driven spin textures such as domain walls and skyrmions~\cite{chen2018compact}. Compared with these approaches, DFL-MTJs fire spontaneously under DC bias, with an intrinsic reset and a voltage-tunable firing rate.

In this work, we focus on the dual-free-layer system and extend the single-free-layer analytical model to describe its dynamics. Building on our previous work on the spiking dynamics of DFL-MTJs~\cite{farcis2023spiking}, we investigate their static and quasi-static properties through a combined experimental and simulation framework. We describe the \emph{windmill} regime and its dependence on device parameters, expanding on the previous single free layer analytical model, with the goal of establishing its boundaries and the design rules to optimize the junction for spiking. Particular attention is given to the role of interlayer coupling, anisotropy distribution, and spin-transfer torque efficiency in determining the magnetization reversal pathways. By systematically varying external stimuli such as bias voltage and applied magnetic field, we quantify the evolution of the energy barrier landscape and characterize the resulting switching thresholds and stability windows. The analysis enables the extraction of the critical device parameters governing the onset of reversal, metastable state formation, and energy dissipation. These results provide a rigorous basis for optimizing DFL-MTJ geometries and material stacks toward controlled, repeatable spiking behavior, which is essential for scalable neuromorphic circuit integration.

\section{Experiments}

The DFL-MTJs structures were deposited by magnetron sputtering onto 100~mm Si wafers. The full multilayer stack consisted of: seed layers / W (0.2~nm) / FeCoB ($t_1$) / MgO (1.1~nm) / FeCoB ($t_2$) / W (0.2~nm) /SAF layer /capping layer, where the numbers in parentheses denote the nominal layer thicknesses. The FeCoB layers were deposited with orthogonal wedge profiles across the wafer, such that the bottom free-layer thickness ($t_1$) varied from 0.6 to 1.0~nm, while the top free-layer thickness ($t_2$) varied from 1.0 to 1.6~nm. The SAF layer at the top of the stack is incorporated to minimize dipolar coupling between the free layers. It consists of the multilayer structure 
Co(0.21)/Ru(0.55)/Co(0.21)/W(0.3)/FeCoB(0.6)/MgO capping (thicknesses in nanometers). The two Co layers are antiferromagnetically coupled via Ruderman–Kittel–Kasuya–Yosida (RKKY) exchange interaction mediated through the Ru spacer layer.

\begin{figure*}[!htbp]
    \centering
    \includegraphics[width=\textwidth,
    clip]{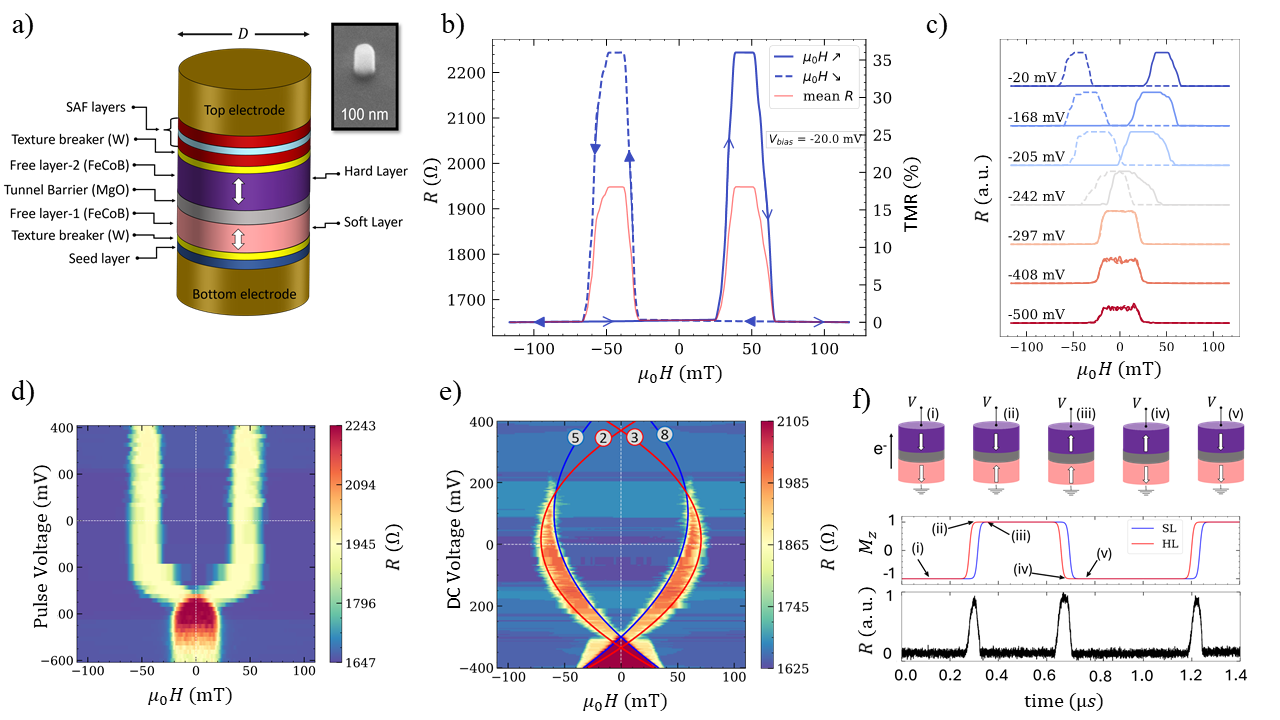}
    \caption{(a) Schematic illustration of a dual free-layer magnetic tunnel junction (DF-MTJ) and a scanning electron microscopy (SEM) image of a fabricated nanopillar with a diameter of 100~nm.
 (b) Representative $R$--$H$ characteristics of the DFL-MTJ measured using 100-ns voltage pulses with an amplitude of $-20$~mV, for increasing (solid blue curve) and decreasing (dashed blue curve) magnetic-field sweeps. The mean $R$--$H$ response of the two bistable states is shown by the solid red curve. 
(c) Evolution of the $R$--$H$ characteristics as a function of the applied pulse voltage. The curves are vertically offset for clarity.
(d) Quasi-static state diagram of the DFL-MTJ as a function of applied magnetic field and pulse-bias voltage, extracted from the mean $R$--$H$ response of the forward and backward field sweeps acquired using 100-ns voltage pulses. 
(e) Quasi-static state diagram of the DFL-MTJ as a function of applied magnetic field and DC bias voltage, extracted from the corresponding $R$--$H$ measurements.
(f) Dual free-layer magnetization dynamic under spin-transfer torque. The switching between four states occurs sequentially, beginning with reflection-electron STT (i$\rightarrow$ii), followed by direct STT (ii$\rightarrow$iii), and then reflection STT (iii$\rightarrow$iv). The cycle repeats, consequently giving rise to characteristic spiking electrical signal. The experimentally obtained resistance response, resulting from toggle switching of the soft (SL) and hard (HL) magnetic layers, is shown together with schematic representations of the time evolution of the magnetization states.\
}
    \label{fig:mtj_stack}
\end{figure*}

Following deposition, the wafers were annealed at 300~$^\circ$C for 10~min under high vacuum (approximately $5 \times 10^{-6}$~mbar) to promote crystallization of the MgO barrier and to enhance the perpendicular magnetic anisotropy (PMA) of the FeCoB layers. The resulting resistance–area (RA) product was approximately 10~$\Omega{\cdot}\mu$m$^2$. 

anopillar MTJ devices with nominal diameters ranging from
50 nm to 100 nm were subsequently defined by electron-beam
lithography, ion-beam etching, and standard photolithography, followed by metallization to form electrical contact pads. A schematic of the wafer stack and a representative scanning electron microscope (SEM) image of a nanopillar device are shown in  Figure~\ref{fig:mtj_stack}a.

The quasi-static magnetoresistance (MR) characteristics of the DFL-MTJs were measured using two complementary schemes. In the first approach, a constant bias voltage was applied across the junction while sweeping the magnetic field perpendicular to the film plane. The resulting TMR typically ranged between 20\% and 40\% for these devices. 
To minimize Joule heating effects, a second measurement protocol was employed in which voltage pulses with a duration of 100~ns were applied across the junction, and the quasi-static resistance states were recorded immediately after each pulse excitation. Furthermore, the real-time spiking dynamics of the DFL-MTJs were investigated using a high-definition oscilloscope (500 MHz analog bandwidth, 2.5 GS/s sampling rate, and 12-bit vertical resolution), enabling high-fidelity acquisition of the transient spiking response (See experimental details in \textcolor{blue}{Supplementary Information}).

The spin-transfer torque (STT) in a double-free-layer magnetic tunnel junction (DF-MTJ) arises from two primary mechanisms: the \emph{reflected-electron STT}, originating from minority carriers reflected at the FeCoB and MgO interface, and the \emph{direct STT}, generated by majority carriers transmitted through the MgO barrier. The manifestation of these effects strongly depends on the relative magnetic orientation of the two free layers. An illustration of this process is presented in Figure~\ref{fig:mtj_stack}f. When both free layers are initially aligned in parallel (P) state, the reflected-electron torque first acts on the layer from which the electrons are injected, driving it toward an antiparallel (AP) configuration (transition i~$\rightarrow$~ii). As the current injection continues, the direct STT becomes dominant, switching the second free layer and restoring the parallel alignment (ii~$\rightarrow$~iii). In this new configuration, the reflected-electron torque once again acts on the injecting layer, reversing its magnetization to the AP state (iii~$\rightarrow$~iv). Subsequently, the direct STT exerts torque on the opposite layer, switching it back to the P state (iv~$\rightarrow$~v).
This sequential switching leads to a self-sustained cyclic evolution of four distinct magnetic configurations, accompanied by two observable resistance spikes, a phenomenon commonly referred to as \textit{\emph{windmill}} dynamics.

Figure~\ref{fig:mtj_stack}b shows an experimentally measured quasi-static resistance–field (\textit{R--H}) characteristic of the junction.
The measurement was performed using 100-ns voltage pulses with an amplitude of $-20$~mV. Starting from negative magnetic saturation, the field is swept toward positive values (solid blue line), causing the soft layer to switch first and driving the junction into the antiparallel (AP) state. With a further increase in the field, the hard layer subsequently reverses, restoring the saturated parallel (P) configuration. Conversely, when the device is initially saturated at a large positive field and the field is swept toward negative values (dashed blue line), the corresponding switching sequence occurs in the opposite direction. Together, these field-driven transitions produce a well-defined bistable hysteretic response over the complete magnetic-field cycle.
The average of these two cycles is shown by the red curve in Figure~\ref{fig:mtj_stack}b. As the magnitude of the negative bias pulse is increased, the influence of STT becomes increasingly pronounced, progressively reducing the critical switching fields of both the soft and hard magnetic layers. The corresponding \textit{R--H} loops measured at different negative bias voltages (Figure~\ref{fig:mtj_stack}c) reveal a systematic evolution of the field-driven switching behavior. At a critical bias of approximately $-297$~mV, the STT becomes sufficiently strong to favor the AP configuration at zero field. With a further increase in the bias magnitude, the system transitions into a regime dominated by \emph{windmill} dynamics at zero field, in which the magnetic layers undergo sustained sequential switching. Signatures of this dynamical regime are also evident in the quasi-static measurements as oscillatory resistance fluctuations at bias voltages beyond the critical threshold. A representative real-time voltage trace acquired at $-350$~mV and zero field using a high-definition oscilloscope is shown in Figure~\ref{fig:mtj_stack}f, directly resolving the temporal evolution of the \emph{windmill} oscillations. The characteristic oscillation frequencies of such devices typically span from a few kilohertz to approximately 100~MHz, depending on the applied bias, as discussed in detail elsewhere~\cite{farcis2024mram}.
The quasi-static state diagrams summarizing the \textit{R--H} characteristics of the device under pulsed and DC bias conditions are shown in Figure~1d and Figure~1e, respectively. These state diagrams are constructed from the mean \textit{R--H} response of the bistable states obtained from the forward and backward magnetic-field sweeps (\textcolor{blue}{see Supplementary Information}). They provide a comprehensive picture of the bias- and field-dependent magnetic response of the device and, importantly, enable identification of the critical bias at which the \emph{windmill} dynamics become self-sustained at zero external magnetic field. A detailed analysis and modeling of the device response under pulsed and DC bias conditions are presented in the following sections.

\section{Modeling}
Understanding the magnetization dynamics responsible for the observed electrical response requires complementing the experimental measurements with numerical simulations. In this work, each ferromagnetic layer is modeled as a single macrospin with uniform magnetization, despite the relatively large diameter of the MTJ pillars (50–100 nm). While the detailed magnetization reversal pathways may involve non-uniform intermediate states ($e.g.$ magnetic domain wall nucleation and propagation \cite{Meo2017}), and may therefore deviate from the macrospin approximation, these transient processes are beyond the scope of this study. Here we focus on the stability boundaries separating the different magnetic regimes for which the macrospin description remains appropriate. This approach enables an efficient exploration of the MTJ behavior across a wide parameter space ($e.g.$ operating conditions, device parameters), where a full micromagnetic treatment would be computationally prohibitive.

The dynamics of a macrospin obeys the Landau-Lifshitz-Gilbert (LLG) equation \cite{landau1935theory} \cite{gilbert1955damping}, with an additional spin-transfer torque term \cite{slonczewski1996current}. The two coupled equations solved here are the following:

 \begin{align}
	\frac{d\mathbf{m}_{SL}}{dt} &= -\gamma_{0}\,(\mathbf{m}_{SL} \times \mathbf{H}_{\mathrm{eff}}^{SL}) + \alpha_{SL} \left(\mathbf{m}_{SL} \times \frac{d\mathbf{m}_{SL}}{dt}\right) \nonumber\\
	&\quad + \gamma_{0}\, a_{DL}^{SL}(V)\, \big(\mathbf{m}_{SL} \times (\mathbf{m}_{SL} \times \mathbf{m}_{HL})\big)
	\label{eq:SL}\\[6pt]
	\frac{d\mathbf{m}_{HL}}{dt} &= -\gamma_{0}\, (\mathbf{m}_{HL} \times \mathbf{H}_{\mathrm{eff}}^{HL})  + \alpha_{HL} \left(\mathbf{m}_{HL} \times \frac{d\mathbf{m}_{HL}}{dt}\right) \nonumber\\
	&\quad - \gamma_{0}\, a_{DL}^{HL}(V)\, \big(\mathbf{m}_{HL} \times (\mathbf{m}_{HL} \times \mathbf{m}_{SL})\big)
	\label{eq:HL}
\end{align}
where $\gamma_0=\mu_0\gamma$, with $\gamma$ the electron gyromagnetic ratio and $\mu_0$ the permeability of free space. $\mathbf{m_{i}}$ is the unitary vector of the magnetization and $\alpha_{i}$ is the damping constant  of each layer $i=HL, SL$. The effective magnetic field $\mathbf{H^i_{eff}}$ acting on the magnetization includes contributions from interfacial magneto-crystalline anisotropy, dipolar interaction, and the Zeeman interaction. The deterministic field is supplemented by a stochastic thermal field, allowing the system to sample the energy landscape \cite{garciapalaciosLazaro1998} and account for thermally activated processes  (see Equation~\ref{thermal noise}). As described by the coupled equations, the mutual interaction between the two magnetic layers is mediated by the spin-transfer torque. For simplicity, we consider firstly the damping-like component of the spin transfer-torque \cite{soren2008STT}, described by a voltage-dependent coefficient $a_{DL}(V)$ :
\begin{align}
    a_{DL}(V) = \frac{\hbar \eta}{2|e|\mu_0 M_S (R \times A) t} V = T_{0} V
    \label{a_DL}
\end{align}
where $\hbar$ is the reduced Plank constant, $\mu_{0}$ is the vacuum permeability, $e$ is electron charge, $t$ is the thickness of the ferromagnetic layer, $M_s$ is the saturation magnetization of the layer and $R \times A$ is the resistance area product of DFL-MTJ pillar. The spin polarization $\eta$ depends on the TMR ratio $\tau$ of the junction as follows $\eta = \frac{\sqrt{\tau\,(\tau+2)}}{2(\tau+1)}$. The prefactor $T_0$ is introduced to simplify the notation. The field-like component can also be included (see Equation~\ref{STT} in the Appendix). Its cumulative role in the dynamic response of the devices and its signature in the state diagram are addressed later.

The above equations were solved numerically using a robust and efficient integration scheme \cite{farcis2023spiking}, enabling the simulation of state diagrams comprising 128×128 operating points. To accurately account for thermal noise, statistical averaging was performed over at least six independent realizations for each magnetic field sweep.

As illustrated in the experimental state diagram in Figure~\ref{fig:mtj_stack}, several distinct states can be identified, separated by transition boundaries. Each boundary represents a critical condition associated with a transition between different magnetic states. To derive analytical expressions for these critical lines, a stability analysis of the different regimes is performed using the Landau–Lifshitz–Gilbert–Slonczewski (LLGS) equations  (Equation~\ref{eq:SL}, Equation~\ref{eq:HL}), reformulated in terms of complex variables through a transformation analogous to the Holstein–Primakoff transformation \cite{SlavinTiberkevich2009}. Within this framework, the magnetization components ($m_x, m_y, m_z$) are transformed into the power-phase variables ($p$, $\phi$). This representation provides a convenient description of the magnetization dynamics in the vicinity of the equilibrium states $(0,0,+1)$ and $(0,0,-1)$ corresponding to the two stable orientations of the perpendicular magnetic anisotropy (PMA) layers. Near these states, the magnetization dynamics can be described as small-amplitude oscillations around the $z$-axis, resulting from the dominant contribution of the anisotropy field along this direction. Since the power variable $p$ of the complex amplitude is directly related to the orientation of the macrospin ($up$ or $down$ state), the corresponding dynamical equation can be analyzed (see Equation~\ref{c-eq} in Appendix for detail) to determine the stability conditions. This approach enables the derivation of analytical expressions for the critical transition lines associated with each magnetic layer, considering the case of a DC voltage $V$ and a DC applied magnetic field $H_{cr}$ at which the state becomes unstable:
\begin{align}
	H_{cr}^{SL} &= -m_{z}^{HL}\!\left( \frac{a_{DL}^{SL} (V) }{\alpha^{SL}}\right) \nonumber\\
	&\quad - m_{z}^{SL}\!\left( \frac{2 K_{u}^{SL}}{M_S^{SL}} - \delta N_{SL} M_S^{SL}\right)
	\label{eq:analytical_lines_SL}\\
	H_{cr}^{HL} &= m_{z}^{SL}\!\left( \frac{a_{DL}^{HL} (V) }{\alpha^{HL}} \right) \nonumber\\
	&\quad - m_{z}^{HL}\!\left( \frac{2 K_{u}^{HL}}{M_S^{HL}} - \delta N_{HL} M_S^{HL}\right)
	\label{eq:analytical_lines_HL}
\end{align}
where $K_{u}^i$ is the PMA anisotropy constant, $M_{S}^i$ is the saturation magnetization and $\alpha^{i}$ is the damping constant of the layer $i={SL,HL}$. $\delta N = N_{zz} - N_{xx}$ is the shape anisotropy factor, where $N_{zz}$ and $N_{xx}$ denote the out-of-plane and in-plane components, respectively, of the demagnetizing tensor for a magnetic layer modeled as a circular disk.

These analytical expressions are subsequently employed in the following section to analyze and interpret the state diagrams, providing insight into the design principles and optimization strategies for magnetic stack engineering.

\begin{figure*}[!tbh]
    \centering
    \includegraphics[width=0.9\textwidth]{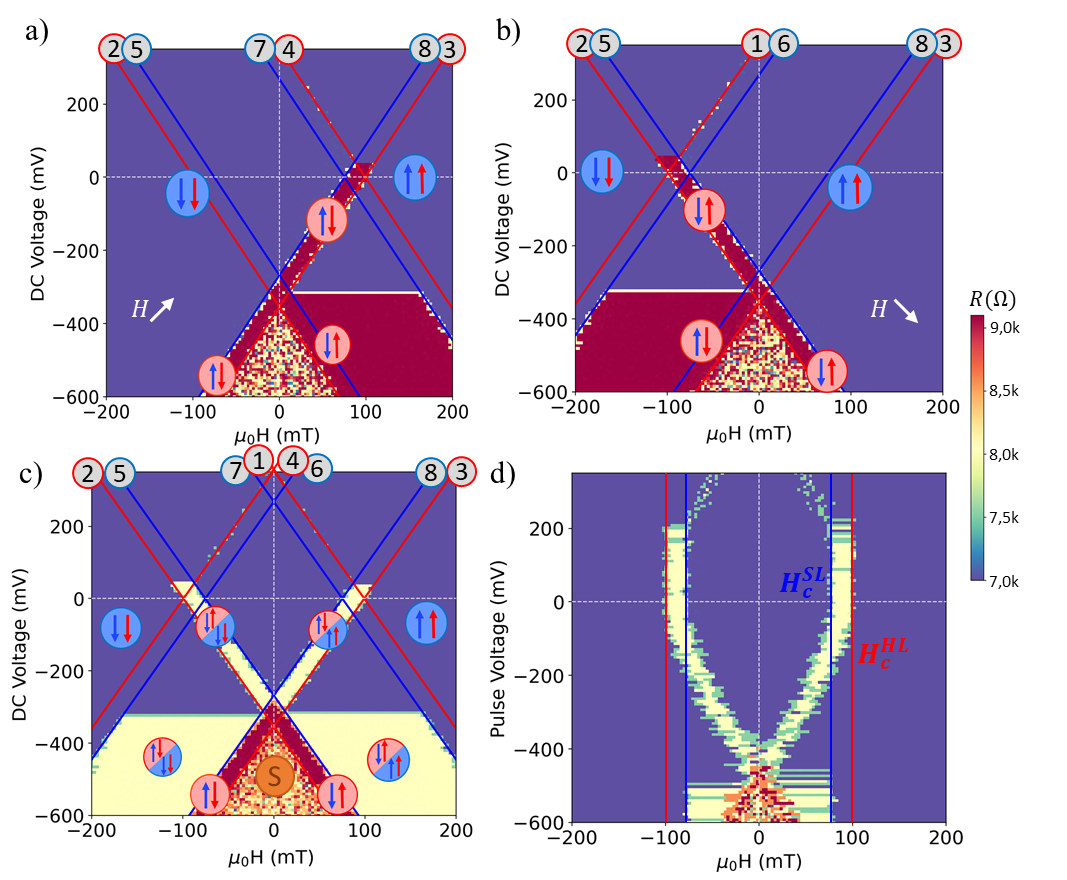}
    \caption{Simulated field--DC voltage state diagrams of a DFL-MTJ for (a) increasing and (b) decreasing applied magnetic field sweeps.
    (c) Superposition of the state diagrams shown in (a) and (b). The blue and red lines
    denote the critical boundaries predicted by the analytical model, with the labels  given in the Table~\ref{tab:lines}. The magnetic states of the system are schematically indicated in each region. (d) Simulated magnetic field–voltage pulse state diagram obtained for a voltage pulse duration of 30~ns. The vertical lines indicate the coercive fields of the soft and hard free layers. The simulation parameters are summarized in Table~\ref{tab:mtj_params}, the electric parameters considered were $TMR = 30\%, R_P=7k\Omega$. The sample temperature was fixed at 300 K. Data points in the DC state diagrams (a)–(c) were averaged over six independent simulation runs, whereas those in the pulsed state diagram (d) were averaged over 18 runs.}
    \label{fig:phase_diag_simul}
\end{figure*}

Figure~\ref{fig:phase_diag_simul} illustrates the numerical procedure used to construct the DFL-MTJ resistance state diagrams. The methodology closely follows the experimental measurement protocol, allowing the system to evolve hysteretically between different magnetic regimes as a function of the applied voltage and external magnetic field. As shown in Figure~\ref{fig:phase_diag_simul}a, the MTJ is initially subjected to a large negative magnetic field, ensuring that both magnetic layers are aligned downward. In this configuration, the MTJ is in the parallel (P) state $down$, which corresponds to the low-resistance state (blue color). A DC bias voltage is applied, and the evolution of the magnetic state is monitored over a duration of 250 ns, which is sufficiently long for the system to reach a steady-state regime. At the end of each time trace, the external magnetic field is incrementally increased (field sweep-up), and the resistance is evaluated again. This procedure is repeated until positive magnetic saturation is reached, where both magnetic layers are aligned upward, corresponding to the parallel (P) state $up$. As the magnetic field and bias voltage are varied, the competition between the different magnetic interactions can drive transitions between distinct magnetic configurations, such as switching from the parallel (P) state to the antiparallel (AP) state or entering the \emph{windmill} dynamic regime (\textcolor{blue}{See  Supplementary Information}). These transitions are reflected by changes in the MTJ resistance, which appear as color variations in the corresponding state diagram.
The procedure is then repeated while sweeping the magnetic field from large positive values to negative saturation, yielding the state diagram shown in Figure~\ref{fig:phase_diag_simul}b, which exhibits different resistive states of the MTJ. 
\begin{table}[!t]
	\centering
	\caption{Parameters of the standard DFL-MTJ.\label{tab:mtj_params}}
	\begin{tabular*}{\columnwidth}{@{\extracolsep\fill}lcc@{\extracolsep\fill}}
		\toprule
		\textbf{Device parameter} & \textbf{Hard Layer} & \textbf{Soft Layer} \\
		\midrule
		Diameter (nm) & 88.6 & 88.6 \\
		Thickness (nm) & 1.1 & 1.1 \\
		$M_s$ (A/m) & $10^6$ & $10^6$ \\
		Damping ($\alpha$) & 0.01 & 0.01 \\
		$K_i$ (J/m$^2$) & $7.06 \times 10^{-4}$ & $6.93 \times 10^{-4}$ \\
		\bottomrule
	\end{tabular*}
\end{table}

By superimposing the field sweep-up and sweep-down results, the complete magnetic field–DC voltage state diagram from Figure~\ref{fig:phase_diag_simul}c is obtained. The bistable regions, shown in yellow, indicate the coexistence of the parallel (P) and antiparallel (AP) states, with the final state determined by the magnetic history of the MTJ. The red regions correspond to the stable AP state, whereas the triangular region exhibiting a checkerboard-like pattern represents the dynamic \emph{windmill} regime (labeled by the letter S from $spiking$), in which the MTJ resistance evolves continuously.

The colormap provides a visualization of the voltage dependence of the different magnetic states; however, its interpretation must be performed with caution due to the degeneracy of the resistance states. Indeed, the parallel configurations with both magnetic layers pointing downward $down-down$ and with both layers pointing upward $up-up$ exhibit the same resistance. Consequently, transitions between these equivalent parallel states do not produce any change in the resistance colormap and cannot be directly identified from the resistance measurement alone.

When the DC voltage is replaced by a voltage pulse, the features of the state diagram are modified, as shown in Figure~\ref{fig:phase_diag_simul}d, in good agreement with the experimental state diagram presented in Figure~\ref{fig:mtj_stack}d. The influence of thermal fluctuations is more pronounced in the pulsed diagram, resulting in a noisier appearance despite the averaging being performed over a larger number of independent simulation runs.

Simulating a complete state diagram is computationally demanding, even within the macrospin approximation, and offers limited flexibility for exploring the parameters space in search of optimal operating conditions. In contrast, the analytical expressions derived above (Eqs. \ref{eq:analytical_lines_SL} and \ref{eq:analytical_lines_HL}) provide an efficient framework for investigating the influence of the different physical parameters on the characteristics of the state diagram. They enable rapid evaluation of how variations in material properties, device geometry, or external conditions modify the transition boundaries, thereby offering valuable guidelines for the optimization and engineering of magnetic stacks.

The comparison between the critical lines predicted by the analytical model and those obtained from numerical simulations, shown in Figure~\ref{fig:phase_diag_simul}(a–c), demonstrates excellent agreement, thereby validating the analytical approach. However, the analysis of the critical lines must be carried out with care, as it is necessary to account for all possible relative orientations of the two coupled magnetic layers and the sweep of the magnetic field.

\begin{table}[!t]
	\centering
	\caption{Critical lines annotated with the first destabilized layer and the orientation of the DFL-MTJ active layers prior to the transition.\label{tab:lines}}
	\begin{tabular*}{\columnwidth}{@{\extracolsep\fill}ccccc@{\extracolsep\fill}}
		\toprule
		\textbf{Number} & \textbf{Layer} & \textbf{HL} & \textbf{SL} & \textbf{Transition} \\
		\midrule
		1 & Hard & up & up & P$\uparrow$ $\rightarrow$ AP$\downarrow$ \\
		2 & Hard & up & down & AP$\uparrow$ $\rightarrow$ P$\downarrow$ \\
		3 & Hard & down & up & AP$\downarrow$ $\rightarrow$ P$\uparrow$ \\
		4 & Hard & down & down & P$\downarrow$ $\rightarrow$ AP$\uparrow$ \\
		\midrule
		5 & Soft & up & up & P$\uparrow$ $\rightarrow$ AP$\uparrow$ \\
		6 & Soft & down & up & AP$\downarrow$ $\rightarrow$ P$\downarrow$ \\
		7 & Soft & up & down & AP$\uparrow$ $\rightarrow$ P$\uparrow$ \\
		8 & Soft & down & down & P$\downarrow$ $\rightarrow$ AP$\downarrow$ \\
		\bottomrule
	\end{tabular*}
\end{table}

As discussed previously, in Figure~\ref{fig:phase_diag_exp} (a) the magnetic field sweep starts from a large negative applied field, for which both magnetic layers are aligned in the $down$ state ($m_{z}^{HL}=-1$; $m_{z}^{SL}=-1$). The two critical lines associated with this configuration are labeled \textbf{4} and \textbf{8} in Table~\ref{tab:lines}. They correspond to the destabilization of either the hard or the soft layer, followed by its reversal to the $up$ state. In the absence of an applied DC voltage, these critical lines reduce to the coercive fields of the hard and soft layers, respectively.

The symmetry of the spin-transfer torque governs the order of magnetic-layer destabilization. In a parallel magnetic configuration, the layer first encountered by the electron flow is destabilized by the spin torque generated by electrons backscattered from the second layer and subsequently spin-polarized. By convention, positive voltage drives electrons through the hard layer first, causing its destabilization prior to the soft layer, whereas negative voltage reverses the switching sequence, with the soft layer switching first.

For a positive voltage, the spin-transfer torque increases the effective damping, thereby favoring a parallel alignment of the two magnetic layers. Consequently, once the hard layer reverses, the soft layer rapidly follows. This behavior is described by critical line \textbf{7}, which separates the antiparallel state $(m_{z}^{HL}=1; m_{z}^{SL}=-1)$ from the parallel state $(m_{z}^{HL}=1; m_{z}^{SL}=1)$. Since this critical line has already been crossed during the field sweep, the reversal of the hard layer from the parallel down state $(m_{z}^{HL}=-1; m_{z}^{SL}=-1)$ to the antiparallel state $(m_{z}^{HL}=1,; m_{z}^{SL}=-1)$ immediately triggers the reversal of the soft layer. As a result, the system undergoes a direct transition from $(m_{z}^{HL}=-1; m_{z}^{SL}=-1)$ to $(m_{z}^{HL}=1; m_{z}^{SL}=1)$, while the two magnetic layers remain effectively aligned throughout the switching process.

For negative voltages, the spin-transfer torque counteracts the intrinsic damping, thereby stabilizing antiparallel configurations. As the magnetic field is swept, the first transition occurs at critical line \textbf{4}, where the soft layer reverses while the hard layer remains in the down state, leading to the antiparallel configuration $(m_{z}^{HL}=-1$; $m_{z}^{SL}=1)$. Upon crossing the critical line associated with the hard layer, the hard layer also reverses, driving the system into the parallel-up state $(m_{z}^{HL}=1; m_{z}^{SL}=1)$ as favored by the Zeeman energy.For larger negative voltages, crossing line \textbf{5} destabilizes this parallel configuration, causing the soft layer to switch back leading to the antiparallel state $(m_{z}^{HL}=1; m_{z}^{SL}=-1)$. As the voltage is further increased in magnitude, crossing line \textbf{2} induces the reversal of the hard layer, returning the system to the parallel-down configuration $(m_{z}^{HL}=-1; m_{z}^{SL}=-1)$. These successive transitions correspond to the sequence of switching events characteristic of the \emph{windmill} dynamics: $\mathrm{AP}{\downarrow} \rightarrow \mathrm{P}{\uparrow} \rightarrow \mathrm{AP}{\uparrow} \rightarrow \mathrm{P}{\downarrow}$. Since critical line \textbf{3} has already been crossed, none of these magnetic configurations remains stable. Instead, each state is destabilized in turn, forcing the system to evolve continuously through the four configurations. The absence of a stable equilibrium gives rise to the \emph{windmill} regime, in which sustained periodic switching of the two magnetic layers results in continuous spiking dynamics. Extending the above analysis to the case where the applied field is swept from positive to negative values (Figure~\ref{fig:phase_diag_simul}b) reveals the two remaining critical lines, $1$ and $6$, which correspond to hysteretic transitions and are symmetric to lines $4$ and $7$, respectively.

\section{Discussion}

\begin{figure*}[!tbh]
    \centering
    \includegraphics[width= 1.0\textwidth]{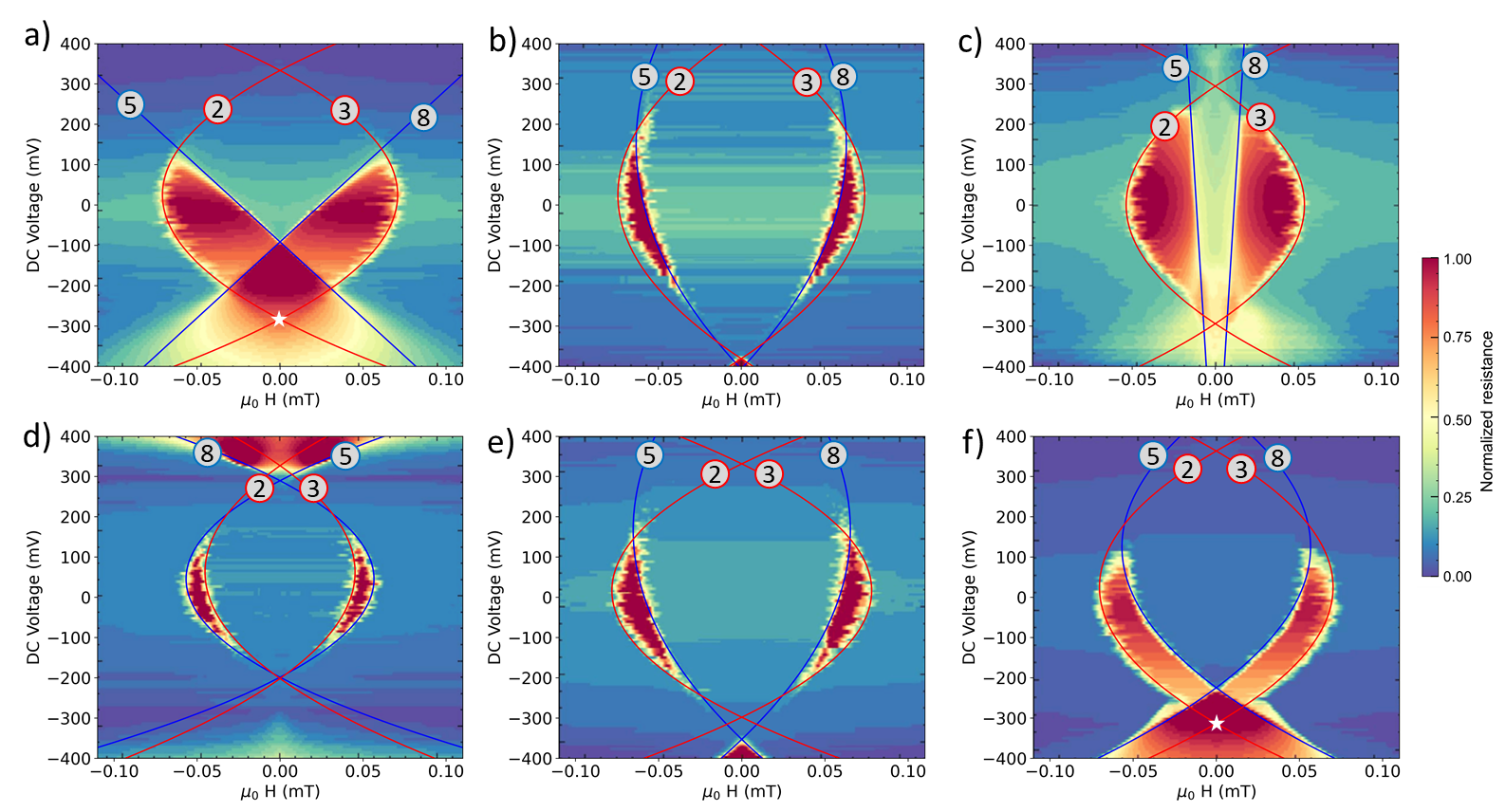}
    \caption{Experimental field–DC voltage state diagrams measured on different devices. The parameters of the devices are listed in the Table~\ref{tab:diagram_params}. For ease of comparison, the resistance of each device is normalized. The blue and red lines denote the critical boundaries predicted by the analytical model using the parameters listed in Table~\ref{tab:diagram_params} and the labeling from the Table~\ref{tab:lines}. The white star in a) and f) indicates the zero-field threshold voltage at which the DFL-MTJ starts spiking.}
    \label{fig:phase_diag_exp}
\end{figure*}

A wide variety of state diagrams has been experimentally measured across the wafer by varying the electrical properties, the lateral dimensions, and the relative thicknesses of the active magnetic layers. Figure~\ref{fig:phase_diag_exp} shows representative examples of these diagrams, where the dynamical regimes discussed above are consistently observed, although their extent, shape, and relative position vary significantly from one device to another.
These results emphasize the complex interplay between the magnetic and electrical properties of the devices and demonstrate the strong impact of the stack parameters on their electrical response. They also highlight the need for a predictive model capable of capturing such a wide range of behaviors while remaining simple.

The analytical critical-line model was applied in its most general formulation (see Equation~\ref{polynom} in the Appendix). The asymmetry of the DFL-MTJ structure is reflected by the voltage-dependent coefficient for both damping-like term and field-like which have polynomial dependence with the voltage \cite{Theodonis2006} \cite{Oh2009} \cite{Boyn2017}. The generic formula used for the critical line are the following:
\begin{align}
	H_{cr}^{SL} &= -m_{z}^{HL}\! \frac{T_{0}^{SL}}{\alpha_{SL}}( \beta_{SL} V + \kappa_{SL} V^2 ) \nonumber\\
	&\quad - m_{z}^{SL}\!\left( \frac{2 K_{u}^{SL}}{M_S^{SL}} - \delta N_{SL} M_S^{SL}\right)
	\label{eq:analytical_lines_SL2}\\
	H_{cr}^{HL} &= m_{z}^{SL}\! \frac{T_{0}^{HL}}{\alpha_{HL}}( \beta_{HL} V + \kappa_{HL} V^2 ) \nonumber\\
	&\quad - m_{z}^{HL}\!\left( \frac{2 K_{u}^{HL}}{M_S^{HL}} - \delta N_{HL} M_S^{HL}\right)
	\label{eq:analytical_lines_HL2}
\end{align}

The coefficients $\beta$ and $\kappa$ characterize the linear and quadratic voltage dependence of the STT torque, respectively and account for a general formalism as the one developed in \cite{Oh2009}. This approach successfully reproduces the main features of all experimentally measured state diagrams. The model parameters were individually calibrated for each device and are summarized in Table~~\ref{tab:diagram_params}. This phenomenological model provides an efficient and compact framework for exploring the parameter space, enabling the rapid identification of optimization strategies and design guidelines for magnetic stack engineering.

Depending on the relative thicknesses of the magnetic layers, the state diagram may undergo a reversal, as illustrated in Figure~\ref{fig:phase_diag_exp}d  and Figure~\ref{fig:phase_diag_exp}e. This inversion results from the interchange of the roles of the two active layers in the DFL-MTJ. For DFL-MTJs intended to operate as spiking neurons, the most favorable state diagrams are those shown in Figures~\ref{fig:phase_diag_exp}a and~\ref{fig:phase_diag_exp}f. 
In these cases, the spiking regime extends over a relatively large region of the state diagram while remaining well separated from the static parallel and antiparallel states, facilitating reliable operation over a wider range of bias conditions. This behavior is achieved in devices with a high tunnel magnetoresistance (TMR) ratio and small diameters. A high TMR ratio increases the spin polarization of the current and thereby enhances the efficiency of spin-transfer torque, whereas reducing the device diameter promotes a more uniform magnetization configuration and suppresses competing magnetic states.

Devices with state diagrams similar to that shown in Figure~\ref{fig:phase_diag_exp}c, characterized by a large asymmetry between the magnetic layers, do not exhibit the desired features for spiking operation in the range of practical operating conditions.

To complete the analysis, we account for the fact that MTJs are resistive devices and therefore experience Joule heating when an electric current flows through the pillar. We investigate this effect by incorporating the temperature $T$ dependence of the magnetic material parameters according to the Callen--Callen scaling law, including the saturation magnetization $M_S(T)$ and the uniaxial magnetic anisotropy  $K_u(T)$ \cite{Diagram_JouleHeating_2018}. Using material parameters reported in the literature, the model reproduces a noticeable curvature of the critical lines; however, this effect becomes significant only at large applied voltages (\textcolor{blue}{See Supplementary Information}). In contrast, the DFL-MTJ devices investigated in this work exhibit a pronounced curvature already at low voltages. This observation therefore suggests that the low-voltage curvature is more likely to originate from the quadratic contribution to the spin-transfer torque rather than from Joule-heating-induced variations of the magnetic properties.

\begin{table*}[!t]
	\centering
	\caption{Device parameters corresponding to the diagrams in Figure~\ref{fig:phase_diag_exp}.\label{tab:diagram_params}}
	\begin{tabular*}{\textwidth}{@{\extracolsep\fill}cccccccccccc@{\extracolsep\fill}}
		\toprule
		\multirow{2}{*}{\textbf{Diagram}} & \multirow{2}{*}{\textbf{Diameter (nm)}} & \multirow{2}{*}{\textbf{$R_p$ ($\Omega$)}} & \multicolumn{2}{c}{\textbf{Thickness (nm)}} & \multicolumn{2}{c}{\textbf{$K_i$ ($10^{-4}$ J/m$^2$)}} & \multirow{2}{*}{\textbf{TMR}} & \multicolumn{2}{c}{\textbf{$\beta$}} & \multicolumn{2}{c}{\textbf{$\kappa$ (V$^{-1}$)}} \\
		\cmidrule{4-5}\cmidrule{6-7}\cmidrule{9-10}\cmidrule{11-12}
		& & & Soft & Hard & Soft & Hard & & Soft & Hard & Soft & Hard \\
		\midrule
		a & 71  & 2520 & 1.11 & 0.79 & 6.57 & 4.98 & 22\% & 2.0 & 0.2 & 0.0   & $-4.0$ \\
		b & 81  & 1932 & 1.34 & 0.81 & 8.17 & 5.13 & 8\%  & 0.9 & 0.2 & $-3.0$  & $-4.0$ \\
		c & 136 & 679  & 1.60 & 0.77 & 9.59 & 4.89 & 6\%  & 0.3 & 0.0 & 0.0   & $-6.0$ \\
		d & 170 & 439  & 1.31 & 0.70 & 8.25 & 4.44 & 3\%  & 2.0 & 1.0 & $-22.0$ & $-8.0$ \\
		e & 102 & 1215 & 1.32 & 0.81 & 8.16 & 5.19 & 11\% & 1.0 & 0.2 & $-3.0$  & $-6.0$ \\
		f & 114 & 968  & 1.27 & 0.96 & 7.85 & 6.10 & 22\% & 1.0 & 0.2 & $-4.0$  & $-4.0$ \\
		\bottomrule
	\end{tabular*}
\end{table*}

\begin{figure*}[!tbh]
    \centering
    \includegraphics[width= 1.0\textwidth]{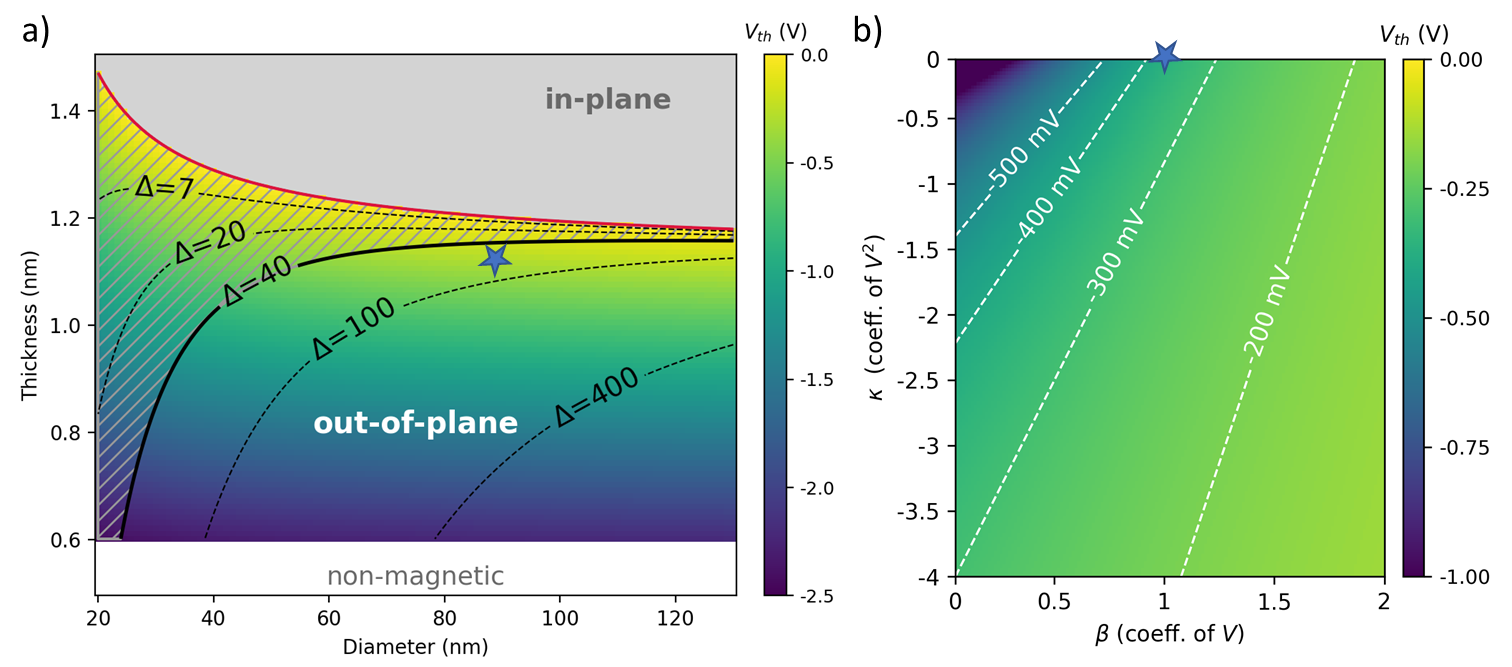}
    \caption{(a) Variation of the zero-field spiking threshold voltage $V_{th}$ with the pillar diameter and thickness of the one active layer, while the thickness of the second layer is kept constant $t = 1.1$~nm. The red line marks the spin-reorientation transition separating the out-of-plane and in-plane states. Black contours indicate the thermal stability factor $\Delta = E_b/k_B T$ at $T = 300$~K. A lower limit for the thickness was imposed, close to the experimentally measured value of the dead-layer thickness. (b) The zero-field spiking threshold voltage $V_{th}$ dependence on the linear ($\beta$) and quadratic ($\kappa$) voltage-torque coefficients (Equation~\ref{eq:analytical_lines_SL2}, Equation~\ref{eq:analytical_lines_HL2}), at given geometry and material parameters. Dashed white lines are iso-$V_{th}$ contours. Parameters used are listed in Table~\ref{tab:mtj_params}. The blue star corresponds to the diagram from Fig~\ref{fig:phase_diag_simul}c.}
\label{vth}
\end{figure*}

A key parameter for the design of spiking MTJ-based neurons is the bias voltage required to trigger the \emph{windmill} dynamics. Figure~\ref{vth} presents the evolution of this zero-field threshold voltage as a function of the device geometrical parameters and electrical transport properties, respectively. For an energy-efficient implementation (see Figure~\ref{vth}a), reducing the lateral dimensions of the pillar and approaching the in-plane/out-of-plane reorientation transition appears to be a favorable strategy. However, this regime also introduces a strong sensitivity to thermal fluctuations, as the magnetic stability factor decreases and can fall below the commonly accepted limit of $\Delta = 40$ at room temperature. Therefore, optimizing the device geometry requires balancing energy efficiency with sufficient thermal stability to ensure reliable operation. 

Another strategy for lowering the zero-field threshold voltage is to tailor the anatomy of the damping-like torque  (Figure~\ref{vth}b). By enhancing the asymmetry between the two magnetic electrodes, the quadratic contribution can be amplified, resulting in a gradual decrease of the voltage required to initiate the \emph{windmill} dynamics.

\section{Conclusion}

Motivated by the potential of dual free-layer magnetic tunnel junctions (MTJs) as artificial neurons, this work investigate the static and dynamic regimes of dual-free-layer magnetic tunnel junctions through a combined experimental and modeling approach. Experimental state diagrams measured under DC and pulsed bias reveal static parallel, antiparallel, and bistable regimes, as well as a dynamical \emph{\emph{windmill}} regime, in which the reversal of one layer triggers the reversal of the other, so that the junction cycles continuously through four magnetic configurations. These diagrams can be reproduced numerically using a macrospin model, allowing a more precise study of the critical lines separating the different regimes. A stability analysis of a projection of the LLG equation onto the anisotropy axis provides an analytical description of these lines and defines the critical voltage at which the \emph{windmill} regime appears. This framework explains the order of layer destabilization set by the symmetry of the spin-transfer torque, and identifies the \emph{windmill} regime as the region where none of the four magnetic configurations remains stable. Combining experiments and simulations also revealed an additional quadratic term in the voltage dependence of the STT, which cannot be explained by Joule heating. This analysis leads to design rules for spiking operation: a high TMR ratio and small pillar diameters favor a wide spiking region, well separated from the static states, whereas a strong asymmetry between the layers is detrimental. The zero-field threshold voltage can be lowered either by approaching the spin-reorientation transition, at the cost of reduced thermal stability, or by enhancing the quadratic torque contribution through electrode asymmetry. This compact analytical description thus provides a practical tool for magnetic stack engineering, and can serve as a basis for circuit-level models of DFL-MTJ neurons for their integration in large-scale spintronic spiking neural networks.

\begin{acknowledgments}

This work was supported in part by the France 2030 government investment plan managed by the French National Research Agency under grant reference PEPR SPIN – SPINTHEORY ANR-22-EXSP-0009, by the French National Research Agency project SpinSpike ANR-20-CE24-0002. The nanofabrication was done on Upstream Technological Platform (PTA). 

\end{acknowledgments}

\section*{Conflicts of Interest}

The authors declare no conflicts of interest.

\newpage
\appendix

\section{}

The effective field $\boldsymbol{H}_{\mathrm{eff}}$ from the LLG equation governing the dynamics of the unit vector of magnetization \textbf{m} is derived from the energy $\epsilon_{\text{tot}}$ as
\begin{align}
    \boldsymbol{H}_{\mathrm{eff}} = -\frac{1}{\mu_0 M_S} \frac{\delta \epsilon_{\mathrm{tot}}}{\delta \boldsymbol{m}},
    \label{Heff}
\end{align}
where $\epsilon_{\text{tot}}$ comprises a magnetocrystalline anisotropy, shape anisotropy, and Zeeman energy,
\begin{align}
\epsilon_K &= K_{u} \left[ 1 - (\boldsymbol{m} \cdot \boldsymbol{u}_k)^2 \right], \label{eq:anisotropy} \\
\epsilon_{\text{dem}} &= \tfrac{1}{2} \mu_0 M_S^2 \left( N_{xx} m_x^2 + N_{yy} m_y^2 + N_{zz} m_z^2 \right), \label{eq:demag} \\
\epsilon_{\text{app}} &= -\mu_0 M_S \boldsymbol{H}_{\text{app}} \cdot \boldsymbol{m}, \label{eq:applied}
\end{align}
where $\boldsymbol{H}_{\text{app}}$ is the external magnetic field, $\boldsymbol{u}_k$ is the magnetocrystalline axis, and $K_{u}$ is the first-order uniaxial anisotropy constant ($K_{u} > 0$). For thin films with interfacial perpendicular magnetic anisotropy (PMA), the anisotropy constant is given by $K_{u}=K_{i}/t_{\mathrm{F}}$, where $K_{i}$ is the interfacial anisotropy and $t_{\mathrm{F}}$ is the film thickness. The diagonal terms of the demagnetizing tensor $(N_{xx}, N_{yy}, N_{zz})$ are evaluated analytically by approximating the magnetic element as an orthorhombic body, following the formalism of \cite{HubertSchafer1998}.

The spin-transfer torque is described by two additional terms in the Landau--Lifshitz--Gilbert (LLG) equation,
\begin{align}
	\left( \frac{d \boldsymbol{m}}{dt} \right)_{\mathrm{STT}} &= -\gamma_0 a_{DL}(V) [ \boldsymbol{m} \times (\boldsymbol{m} \times \boldsymbol{p})] \nonumber\\
	&\quad + \gamma_0 a_{FL}(V) (\boldsymbol{m} \times \boldsymbol{p}),
	\label{STT}
\end{align}
where $\boldsymbol{p}$ denotes the unit vector along the current spin polarization, while $a_{DL}$ and $a_{FL}$ represent the damping-like and field-like torque coefficients, respectively, both of which depend on the applied voltage $V$.

Thermal fluctuations at finite temperature $T$ are accounted for as a Gaussian distributed thermal field $\boldsymbol{H}_{\mathrm{th}}$ added to the effective field $\boldsymbol{H}_{\mathrm{eff}}$ (Equation~\ref{Heff}) \cite{Brown1963}. This fluctuating field has the following properties
\begin{align}
\langle \boldsymbol{H_{\mathrm{th}}}(t) \rangle &= 0, \\
\langle \boldsymbol{H_{\mathrm{th}}}(t)\, \boldsymbol{H_{\mathrm{th}}}(t') \rangle
&= \frac{2 \alpha k_B T}{\gamma \mu_0 M_S \Omega}\,\delta(t-t'),
\label{thermal noise}
\end{align}
where $\delta$ is the Dirac function, $k_{B}$ is the Boltzmann constant and $\Omega$ is the sample volume. The procedure used to generate the white noise follows standard methods \cite{numericalRecipes}.

To derive the critical conditions for switching and instability, the Landau--Lifshitz--Gilbert--Slonczewski (LLGS) equation is recast in the power--phase representation. Introducing the complex amplitude $c = \sqrt{p}e^{-i \phi}$, the magnetization dynamics can be written as
\begin{align}
	\frac{d\mathbf{c}}{dt} &= -i \mathbf{c} \Big( \omega_H - P_j (\omega_{DL}+\alpha \omega_{FL}) \nonumber\\
	&\qquad + (\omega_K + \omega_N)(1 - \mathbf{c} \bar{\mathbf{c}}) \Big) \nonumber \\
	&\quad - \alpha \mathbf{c} (1 - \mathbf{c} \bar{\mathbf{c}}) \Big[ \omega_H + P_j \Big( \frac{\omega_{DL}}{\alpha} - \omega_{FL} \Big) \nonumber\\
	&\qquad + (\omega_K + \omega_N)(2 \mathbf{c} \bar{\mathbf{c}} - 1) \Big].
	\label{c-eq}
\end{align}
$P_{i}$ and $P_{j}$ are unitary values indicating the orientation of the polarizer ($+1$ for \textit{up} and $-1$ for \textit{down}). The angular frequencies are defined as
$\omega_H = \gamma_0' H_{\mathrm{app}}$, $\omega_{DL} = \gamma_0' a_{DL} (V) / \mu_0$, $\omega_{FL} = \gamma_0' a_{FL} (V) / \mu_0$, $\omega_K = \gamma_0' H_K$, and $\omega_N = \gamma_0' (N_{zz} - N_{xx}) M_S$, with $\gamma_0' = \gamma_0 / (1 + \alpha^2)$ a reduced gyromagnetic factor.

Separating the real and imaginary parts of the complex amplitude equation yields the following coupled equations for the power $p$ and phase $\phi$,

\begin{subequations}
	\begin{align}
		\frac{dp}{dt}
		&= -2\alpha(1-p)
		\left[
		\begin{aligned}
			&\omega_H
			+ P_j\left(\frac{\omega_{DL}}{\alpha}-\omega_{FL}\right)\\
			&\quad - (\omega_K+\omega_N)(2p-1)
		\end{aligned}
		\right],
		\\
		\frac{d\phi}{dt}
		&= -
		\left[
		\omega_H
		- P_j(\omega_{DL}+\alpha\omega_{FL})
		+ (\omega_K+\omega_N)(1-p)
		\right].
	\end{align}
\end{subequations}

To identify the stability threshold, and neglecting the FL torque contribution, a linear stability analysis is performed around steady-state, yielding the critical line equation,
\begin{align}
\omega_{H,i} + P_j \left( \frac{\omega_{DL,i}}{\alpha_i}-\omega_{FL,i} \right) + P_i (\omega_{K,i} + \omega_{N,i}) = 0.
\end{align}
This is used for predicting state diagrams in the case of coupled layer $i,j = \mathrm{HL},\mathrm{SL}$. In its most general form, the STT contribution can be written as a polynomial function of the applied voltage $V$ \cite{Theodonis2006,Oh2009,Boyn2017},
\begin{align}
f(V) = \frac{\omega_{DL,i}}{\alpha_i}-\omega_{FL,i} = a_0 + a_1 V + a_2 V^2+\dots,
\label{polynom}
\end{align}
where $a_i$ are the coefficients of the polynomial expansion.

\pagebreak
\clearpage
\begin{widetext}
	\begin{center}
		\textbf{\large Supplementary Information}
	\end{center}
	\setcounter{secnumdepth}{0}
	\setcounter{secnumdepth}{0}
	\setcounter{equation}{0} \setcounter{figure}{0} 
	\setcounter{table}{0} %
	\renewcommand{\theequation}{S\arabic{equation}}
	\renewcommand{\thefigure}{S\arabic{figure}}
	\renewcommand{\thetable}{S\Roman{table}}

\section{Experimental methodology and state diagram}
To optimize the magnetic properties of the dual-free-layer magnetic tunnel junction (DFL-MTJ), a combinatorial deposition strategy was employed during sputter growth. Orthogonal thickness wedges were introduced across the wafer, producing a systematic variation in the thicknesses of the two free magnetic layers. The thicknesses of the two FeCoB free layers were systematically varied using orthogonal thickness wedges: the bottom free layer ($t_1$) along the ($x$)-direction from approximately 0.6 to 1.0 nm, and the top free layer ($t_2$) along the ($y$)-direction from approximately 1.0 to 1.6 nm.
This approach enabled the exploration of a broad parameter space within a single deposition run, providing an efficient means of identifying the optimal composition of the DFL stack.

Following deposition, the magnetic properties of the unpatterned wafer were evaluated by magneto-optical Kerr effect (MOKE) magnetometry with the magnetic field applied perpendicular to the film plane. The resulting coercivity map, shown in Figure~\ref{sfig:exp_map}(a), directly reflects the influence of the free-layer thicknesses on the perpendicular magnetic anisotropy and switching characteristics of the stack. The continuous variation of coercivity across the wafer confirms the effectiveness of the wedge deposition and provides guidance for selecting the most suitable layer thickness combination for device fabrication.

\begin{figure*}[!tbh]
	\centering
	\includegraphics[width= 1.0\textwidth]{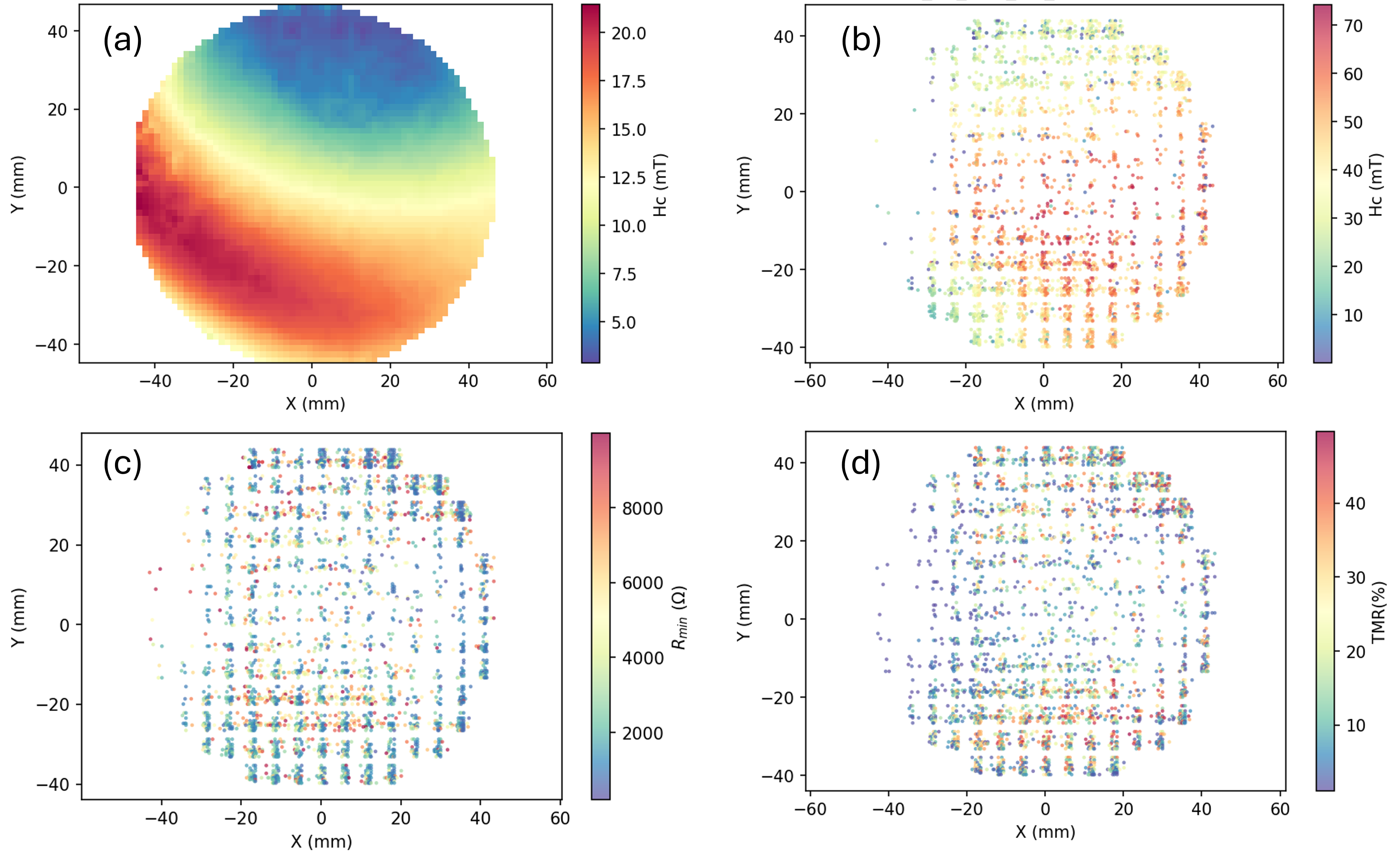}
	\caption{(a) Coercivity map of the unpatterned DFL stack (blank wafer) measured by magneto-optical Kerr effect (MOKE) magnetometry with the magnetic field applied perpendicular to the film plane. (b) Coercivity map of the patterned DFL-MTJ devices across the entire wafer, measured with the magnetic field applied perpendicular to the film plane. (c) Resistance $(R_{\min} = R_{\mathrm{P}})$ map of the patterned DFL-MTJ devices across the wafer.(d) Tunnel magnetoresistance (TMR) map of the patterned DFL-MTJ devices across the wafer.}
	\label{sfig:exp_map}
\end{figure*}

The wafer was subsequently patterned into circular DFL-MTJ devices with nominal diameters of 50, 80, and 100 nm using the standard fabrication process described above. Electrical characterization was performed using an automated wafer-probing station equipped with a perpendicular electromagnet, enabling high-throughput measurements of every accessible device across the wafer. For each MTJ, resistance as a function of perpendicular magnetic field (R--H) was recorded automatically, allowing the extraction of the coercive fields of the individual free layers, the tunnel magnetoresistance (TMR), and the junction resistance.

The wafer-scale distributions of these electrical properties are summarized in Figure~\ref{sfig:exp_map}(b--d). The coercivity map obtained from the patterned devices (Figure~\ref{sfig:exp_map}b) shows that the overall magnetic behavior observed by MOKE is preserved after nanofabrication, while the coercive fields increase systematically compared with those of the blanket film, as expected from the enhanced shape anisotropy associated with lateral confinement at the nanoscale. The map also captures the device-to-device variations arising from lithographic processing and dimensional scaling. 
The resistance and TMR maps (Figure~\ref{sfig:exp_map}c, d) further demonstrate the spatial uniformity and reproducibility of the fabricated DFL-MTJs across the wafer. The resistance map corresponds to the minimum junction resistance, ($R_{\min}$), measured in the parallel magnetic configuration, ($R_{\mathrm{P}}$). To ensure the statistical reliability of the extracted device parameters, each resistance--field (\textit{R--H}) loop was measured 15 consecutive times over the full magnetic-field sweep. The final \textit{R--H} characteristic used for analysis was obtained by averaging these repeated measurements, thereby reducing measurement noise and improving the precision of the extracted resistance, TMR, and coercive fields.

Together, these wafer-level magnetic and electrical characterizations establish the relationship between the deposited free-layer thicknesses, the resulting magnetic properties, and the performance of the fabricated DFL-MTJ devices, thereby identifying the optimal region of the wafer for the detailed investigations presented in the this study.

\begin{figure*}[!tbh]
	\centering
	\includegraphics[width= 1.0\textwidth]{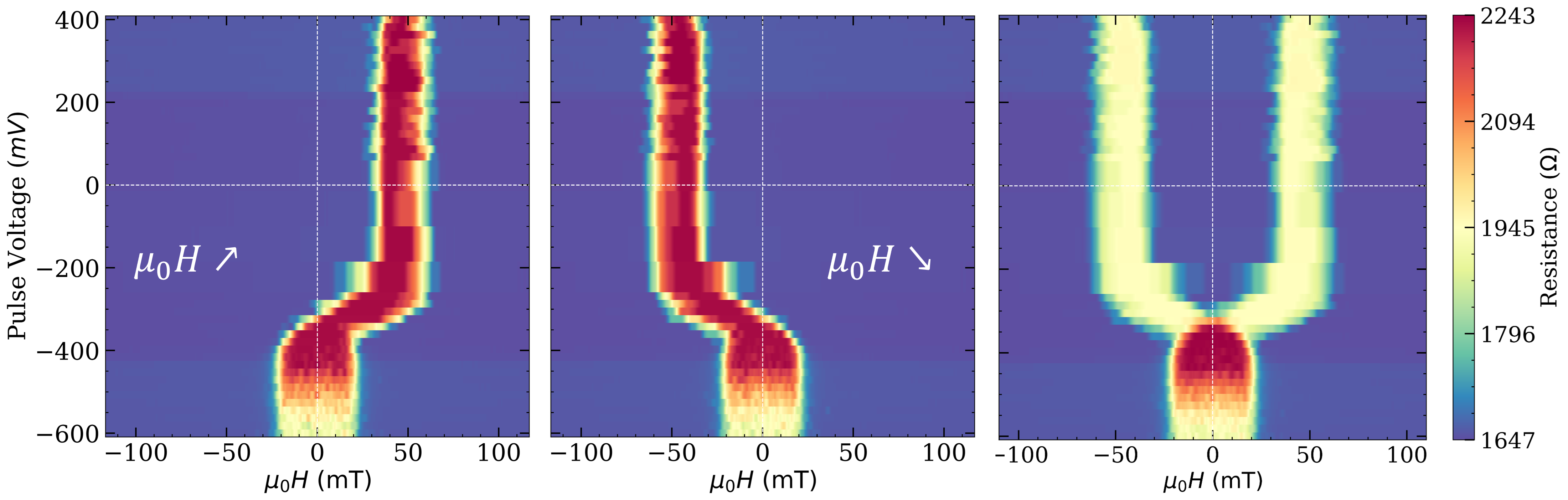}
	\caption{Experimental state diagrams measured with pulse-bias voltage when the applied magnetic field a) increases, b) decreases and c) the mean of the previous 2 measurements. This methodology corresponds to the Figure1.d  from the main text.}
	\label{sfig:phase_diag_exp}
\end{figure*}

To investigate the quasi-static magnetic response of the DFL-MTJ under electrical excitation, we experimentally mapped the device resistance as a function of applied magnetic field and pulse-bias voltage. Figure~\ref{sfig:phase_diag_exp} presents a quasi-static state diagram obtained using voltage pulses with a duration of 100 ns.
For each pulse-bias amplitude, a complete \textit{R--H} loop was measured 15 times over the full magnetic-field sweep. The final \textit{R--H} characteristic was obtained by averaging these repeated measurements, providing a robust and statistically reliable representation of the device response and enabling the evolution of the resistance states to be tracked over a wide range of bias conditions.

Figures~\ref{sfig:phase_diag_exp}a and~\ref{sfig:phase_diag_exp}b show the state diagrams obtained during increasing and decreasing magnetic-field sweeps, respectively. Owing to the hysteretic nature of the DFL-MTJ, the two sweep directions exhibit different switching trajectories, reflecting the sequential reversal of the soft and hard free layers. To provide a more compact representation of the device behavior, Figure~\ref{sfig:phase_diag_exp}c presents the mean response obtained by averaging the forward and backward field sweeps.

The resulting state diagram clearly reveals the bias dependence of the critical switching fields of the two magnetic layers. As the magnitude of the negative pulse bias voltage increases, spin-transfer torque progressively reduces the switching fields, causing the stability regions of the parallel and antiparallel states to shrink. Above a critical bias, the bistable magnetic states collapse into a region dominated by self-sustained windmill dynamics, which appears in the state diagram as a distinct dynamical regime separating the two static magnetic configurations. Consequently, the quasi-static state diagram provides a convenient visualization of the evolution of the magnetic states under combined magnetic field and electrical excitation while directly identifying the onset of the windmill dynamic regime.

\section{Windmill dynamics}

Although the main text provides a detailed discussion of the state diagrams, it does not include representative numerical time traces. Here, we present example simulated time traces of the windmill dynamics. As shown in Figure~\ref{sfig:supp_windmill}b--d,  the device remains in the antiparallel (AP) state for most of the time and undergoes rapid magnetization reversals when transitioning between the two degenerate AP configurations ($up$--$down$ and $down$--$up$). The resistance exhibits sharp dips corresponding to these rapid transitions. As the applied voltage increases, the spiking frequency also increases, illustrating the voltage tunability of the windmill dynamics.
\begin{figure*}[!tbh]
	\centering
	\includegraphics[width= 1.0\textwidth]{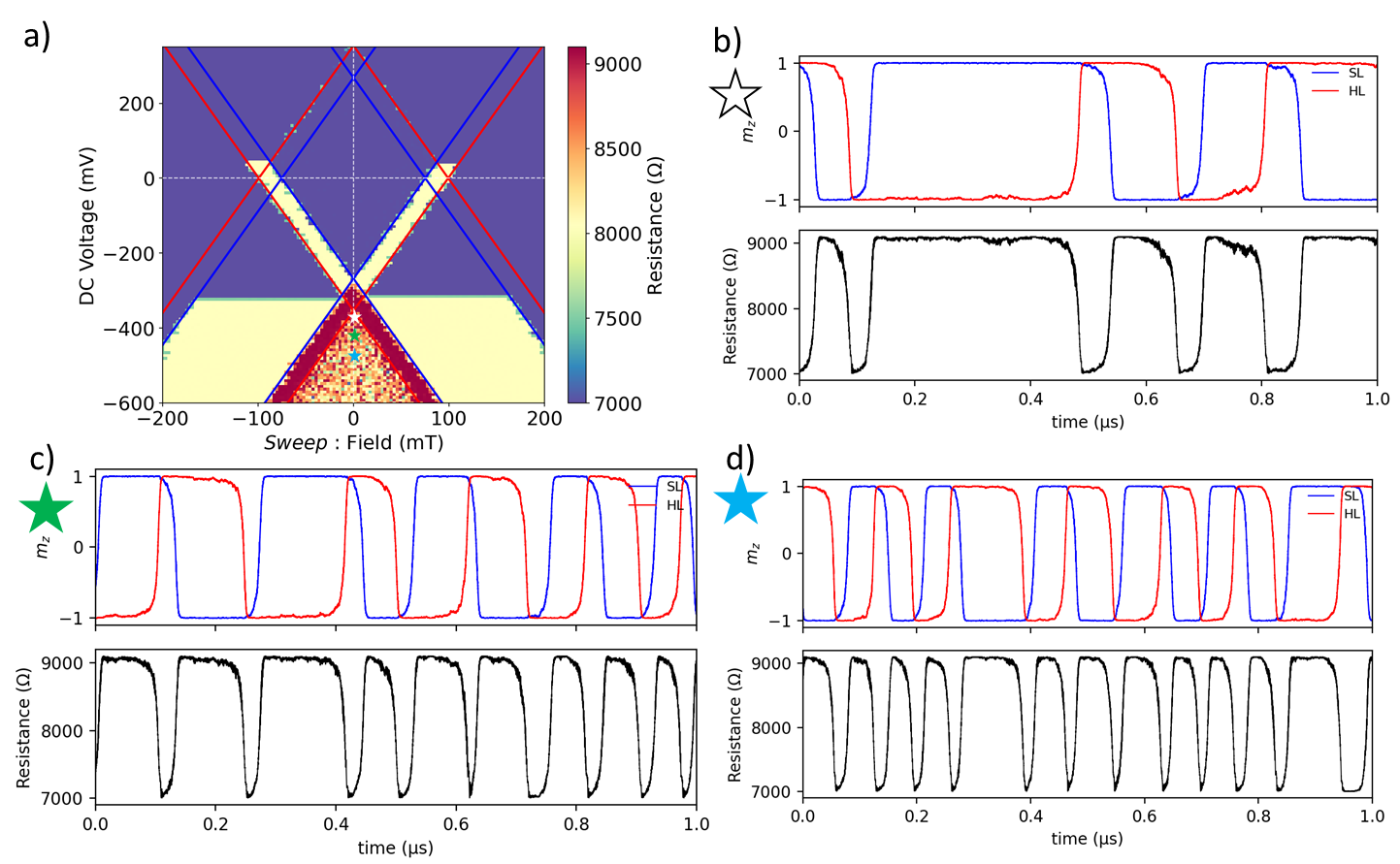}
	\caption{(a) Simulated field--DC voltage state diagram of a DFL-MTJ, identical to Figure~2(c) of the main text. The critical boundaries predicted by the analytical model are shown in red (hard layer) and blue (soft layer). The white, green, and cyan stars indicate three operating points within the windmill regime at zero applied field and $V=-350$, $-400$, and $-450$~mV, respectively. (b--d) Simulated time traces of the soft-layer (SL) and hard-layer (HL) magnetization, $m_z$ (top), together with the corresponding device resistance (bottom), for the operating points indicated by the white (b), green (c), and cyan (d) stars in panel (a).  
	}
	\label{sfig:supp_windmill}
\end{figure*}

\section{Thermal effect and Joule heating}

In experiments, Joule heating is intrinsically present because of the electric current flowing in the device and must be taken into account. Temperature affects the device through both the temperature dependence of the material parameters and the thermal fluctuations it induces. Before addressing these temperature-dependent effects, we first estimate the dependence of the MTJ temperature on the applied bias voltage using a one-dimensional thermal model derived from the first law of thermodynamics. The thermal dynamics of the MTJ are described by:
\begin{align}
	C\frac{dT}{dt} + Q(T - T_0)= \frac{V^2}{R} 
\end{align}
where $\frac{V^2}{R}$ is the Joule heating power, $C$ is the heat capacity o the MTJ, $R$ is its electrical resistance, and $Q$ is the heat transfer coefficient between the tunnel barrier and the conductive electrodes which are assumed to remain at room temperature $T_0$. Under a constant DC bias, the device temperature increases until it reaches a steady-state value given by:
\begin{align}
	T= T_0 + \frac{V^2}{QR}
\end{align}
In the first approach, the device temperature  $T$ directly influences the thermal fluctuations, which are introduced in the model through a Gaussian-distributed thermal field $\boldsymbol{H_{th}}$ added to the effective field $\boldsymbol{H_{eff}}$. This fluctuating field is characterized by the following properties:
\begin{align}
	\langle \boldsymbol{H_{\mathrm{th}}}(t) \rangle &= 0, \\
	\langle \boldsymbol{H_{\mathrm{th}}}(t)\, \boldsymbol{H_{\mathrm{th}}}(t') \rangle
	&= \frac{2 \alpha k_B T}{\gamma \mu_0 M_S \Omega}\,\delta(t-t').
	\label{seq:thermal noise}
\end{align}
where $\delta$ is the Dirac function, $k_{B}$ is the Boltzmann constant, $\alpha$ is the damping constant, $\gamma$ is the gyromagnetic ratio, $\mu_0$ is the vacuum permeability, $M_S$ the saturation magnetization and $\Omega$ is the sample volume. The procedure used to generate the white noise follows standard methods \cite{numericalRecipes}.

In a second approach, the influence of temperature T on the intrinsic material properties is considered. The temperature dependence of the saturation magnetization $M_s(T)$, anisotropy constants $K_u(T)$ and $K_i(T)$, and spin polarization $\eta$ is introduced through the Callen--Callen scaling laws \cite{Diagram_JouleHeating_2018}. In particular, the reduction of $\eta$ with temperature directly affects the spin-transfer torque amplitude through its role as a prefactor of $a_{DL}$.

\begin{align}
	\frac{M_s(T)}{M_s(0)} &= \left[1-\left(\frac{T}{T_C}\right)^{a}\right]^{b}\\
	\frac{K_u(T)}{K(0)} &= \left [\frac{M_s(T)}{M_s(0)}\right]^{c}\\
	P(T) &= P_0 \left[1-\alpha_P\left(\frac{T}{T_b}\right)\right]^{3/2}\\
	\mathrm{TMR}(T) &= \frac{2P(T)^2}{1-P(T)^2}\\
	\eta(T) & = \frac{\sqrt{TMR(T)\,(TMR(T)+2)}}{2(TMR(T)+1)}\\
	a_{DL}(T)  &= \frac{\hbar \eta (T)}{2|e|\mu_0 M_S(T) (R\times A) t}
\end{align}

\noindent where $T_C$ is the Curie temperature and $a$, $b$, $c$ are material-dependent scaling exponents. The temperature-renormalized $M_S(T)$ and $K(T)$ are then used to evaluate the field and energy terms, as well as for the critical lines. Regarding transport parameters, $T_b$ is the blocking temperature for the spin polarization, $P(T)$ is the spin polarization, and $\eta$ is the spin transfer torque efficiency. $\eta$ and $M_s(T)$ are then used to evaluate the STT term $a_{DL}$.

Where for each value of voltage not only the spin-transfer torque term but every material parameter have a slight shift. This slowly lead the lines away from the linear model as seen in Figure~\ref{sfig:joule_heating_effect}.

Figure~\ref{sfig:joule_heating_effect}a show the representative junction from main text, where only the damping like linear term is considered, thus making critical lines linear. In Figure~\ref{sfig:joule_heating_effect}b is show the results for the critical line model for three values of the effective thermal conductance $Q$ entering
$T(V) = V^2/(RQ) + T_0$, thus changing the sample temperature.
The lines remain nearly straight near $V = 0$ in
all three cases and only bend away from zero bias, showing that
Joule heating cannot account for the low-voltage curvature observed experimentally in the state diagrams.

\begin{figure*}[!tbh]
	\centering
	\includegraphics[width= 1.0\textwidth]{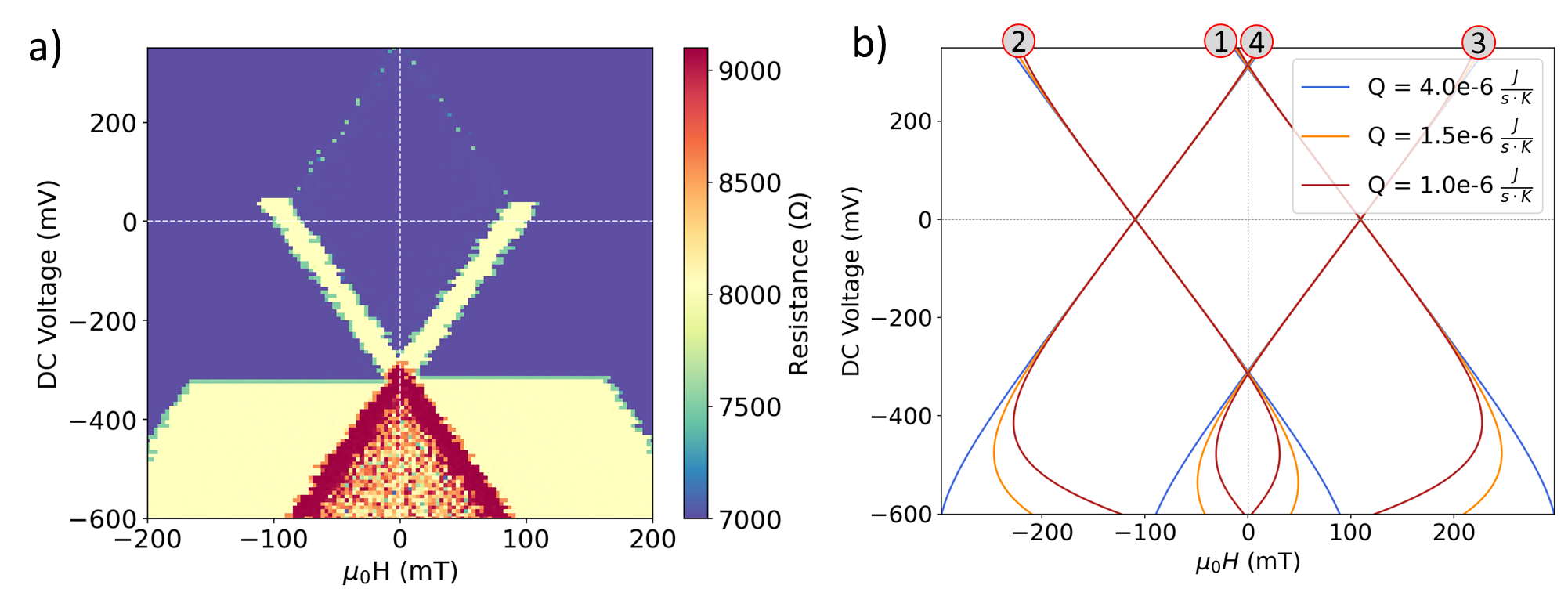}
	\caption{a) Numerical DC  state diagram of the representative junction cited in the main text Table I, without quadratic term in STT ($\beta = 1 ; \kappa=0)$. (b) Model of  critical
		lines obtained from the Joule-heating model alone ($\kappa = 0$).}
	\label{sfig:joule_heating_effect}
\end{figure*}
\begin{table}[H]
	\centering
	\caption{Material and Joule-heating parameters used for the hard layer in the critical-line simulations shown in Figure~\ref{sfig:joule_heating_effect}.}
	\label{stab:heating_params}
	\small
	\begin{tabular}{|l|c|}
		\hline
		\textbf{Parameter} & \textbf{Value} \\
		\hline
		$M_s$      & $1.0\times10^6\,\mathrm{A/m}$ \\
		Diameter   & $88.6\,\mathrm{nm}$ \\
		Thickness  & $1.1\,\mathrm{nm}$ \\
		$K_i$      & $7.095\times10^{-4}\,\mathrm{J/m^2}$ \\
		$\alpha$   & 0.01 \\
		$R_p$      & $7000\,\Omega$ \\
		TMR        & 30\% \\
		\hline
	\end{tabular}
	\hspace{1.5cm}
	\raisebox{-0.42\height}{
		\begin{tabular}{|l|c|}
			\hline
			\multicolumn{2}{|c|}{\textbf{Callen--Callen}} \\
			\hline
			$(a,b,c)$ & $(1.73,1.0,2.0)$ \\
			$T_C$       & 800 K \\
			$T_0$       & 300 K \\
			\hline
	\end{tabular}}
\end{table}


\section{Simulated state diagrams }

While only numerical diagrams with linear spin-transfer-torque terms are shown in the main text, it is possible to include quadratic terms in the numerical simulations to obtain diagrams that closely resemble the experimental ones. This is achieved using the same material parameters as in experiment, with $K_i$ extracted from the coercive fields, and $\beta$
and $\kappa$ extracted from the experimental state diagrams as shown in Table~\ref{stab:diagram_params}.
\begin{figure*}[!tbh]
	\centering
	\includegraphics[width= 1.0\textwidth]{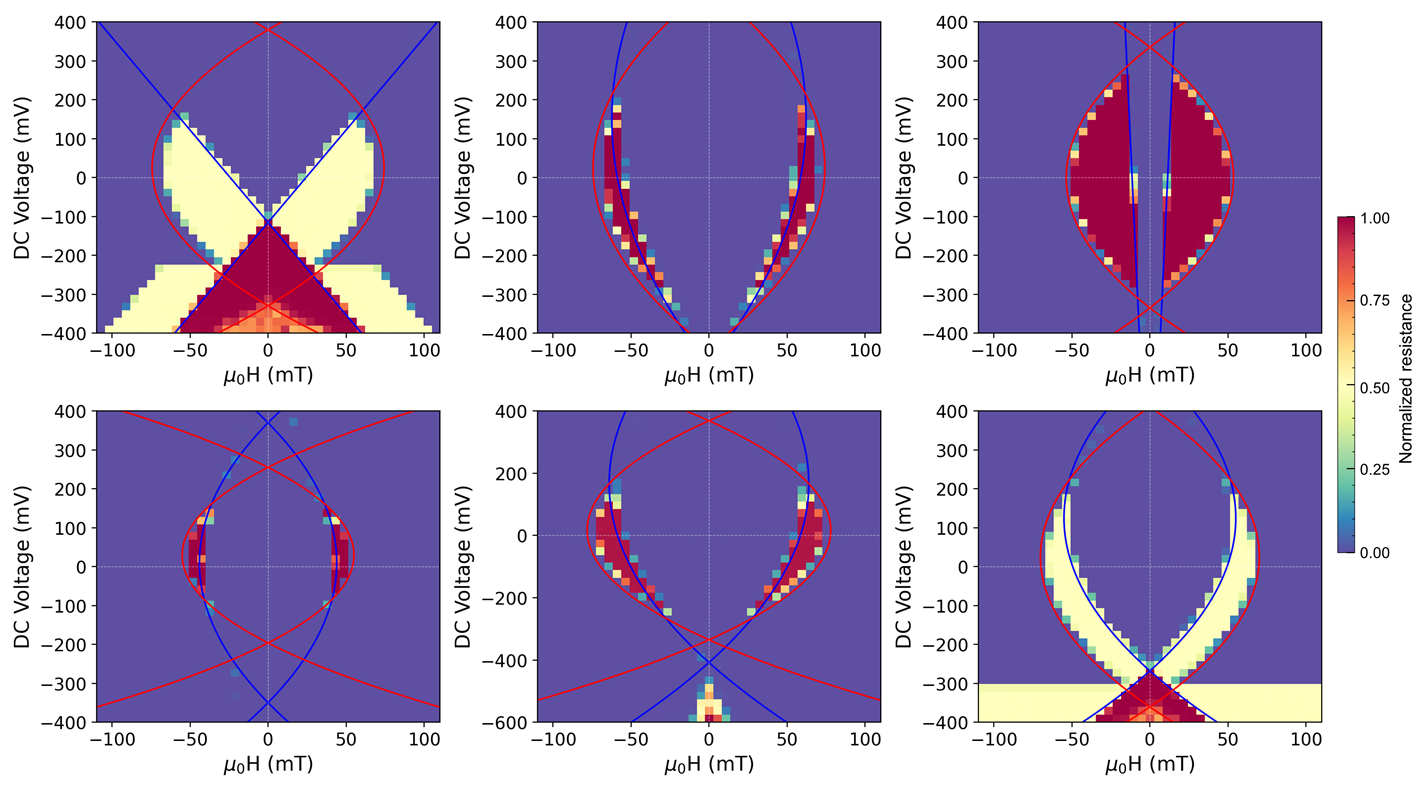}
	\caption{Field–DC voltage state diagrams simulated for different devices. The parameters of the devices are listed in the Table~\ref{stab:diagram_params}. For ease of comparison, the resistance of each device is normalized. The blue and red lines denote the critical boundaries predicted by the analytical model using the parameters listed in Table~\ref{stab:diagram_params} and the labeling from the Table II in main. }
	\label{sfig:phase_diag_exp_lines}
\end{figure*}
\begin{table}[t]
	\centering
	\caption{Device parameters corresponding to the diagrams in Figure~\ref{sfig:phase_diag_exp}.}
	\label{stab:diagram_params}
	\small
	\renewcommand{\arraystretch}{1.25}
	
	\begin{tabular*}{\textwidth}{@{\extracolsep{\fill}}cccccccccccc@{}}
		\toprule
		\multirow{2}{*}{\textbf{Diagram}}
		& \multirow{2}{*}{\textbf{Diameter}}
		& \multirow{2}{*}{$\boldsymbol{R_p}$}
		& \multicolumn{2}{c}{\textbf{Thickness (nm)}}
		& \multicolumn{2}{c}{\boldsymbol{$K_i$} (\textbf{$10^{-4}$ J/m$^2$})}
		& \multirow{2}{*}{\textbf{TMR}}
		& \multicolumn{2}{c}{$\boldsymbol{\beta}$}
		& \multicolumn{2}{c}{$\boldsymbol{\kappa}$ (V$^{-1}$)} \\
		
		\cmidrule(lr){4-5}
		\cmidrule(lr){6-7}
		\cmidrule(lr){9-10}
		\cmidrule(lr){11-12}
		
		& \textbf{(nm)}
		& \textbf{($\Omega$)}
		& \textbf{Soft} & \textbf{Hard}
		& \textbf{Soft} & \textbf{Hard}
		&
		& \textbf{Soft} & \textbf{Hard}
		& \textbf{Soft} & \textbf{Hard} \\
		
		\midrule
		a & 71  & 2520 & 1.11 & 0.79 & 6.57 & 4.98 & 22\% & 2.0 & 0.2 & 0.0    & $-4.0$ \\
		b & 81  & 1932 & 1.34 & 0.81 & 8.17 & 5.13 & 8\%  & 0.9 & 0.2 & $-3.0$ & $-4.0$ \\
		c & 136 & 679  & 1.60 & 0.77 & 9.59 & 4.89 & 6\%  & 0.3 & 0.0 & 0.0    & $-6.0$ \\
		d & 170 & 439  & 1.31 & 0.70 & 8.25 & 4.44 & 3\%  & 2.0 & 1.0 & $-22.0$ & $-8.0$ \\
		e & 102 & 1215 & 1.32 & 0.81 & 8.16 & 5.19 & 11\% & 1.0 & 0.2 & $-3.0$ & $-6.0$ \\
		f & 114 & 968  & 1.27 & 0.96 & 7.85 & 6.10 & 22\% & 1.0 & 0.2 & $-4.0$ & $-4.0$ \\
		\bottomrule
	\end{tabular*}
\end{table}

\end{widetext}
\pagebreak
\bibliography{apssamp}

\end{document}